# Tungsten disulfide-gold nanohole hybrid metasurfaces for nonlinear metalens in the visible region

*Jiawei Chen[1], Kai Wang[1,*], Hua Long[1], Xiaobo Han[2], Hongbo Hu[1], Weiwei Liu[1], Bing Wang[1,*], Peixiang Lu[1,2,*]*

[1]Wuhan National Laboratory for Optoelectronics and School of Physics, Huazhong University of Science and Technology, Wuhan 430074, China

[2]Laboratory of Optical Information Technology, Wuhan Institute of Technology, Wuhan 430205, China

*Corresponding authors: kale_wong@hust.edu.cn (KW), wangbing@hust.edu.cn (BW), lupeixiang@hust.edu.cn (PXL)

**Abstract**

Recently, nonlinear hybrid metasurface comes into an attractive new concept in the research of nanophotonics and nanotechnology. It is composed of semiconductors with an intrinsically large nonlinear susceptibility and traditional plasmonic metasurfaces, offering opportunities for efficiently generating and manipulating nonlinear optical responses. A high second-harmonic generation (SHG) conversion efficiency has been demonstrated in the mid-infrared region by using multi-quantum-well (MQW) based plasmonic metasurfaces. However, it has yet to be demonstrated in the visible region. Here we present a new type of nonlinear hybrid metasurfaces for the visible region, which consists of a single layer of tungsten disulfide ($WS_2$) and a phased gold nanohole array. The results indicate that a large SHG susceptibility of $\sim 10^{-1}$ nm/V at 810 nm is achieved, which is 2~3 orders of magnitude larger than that of typical plasmonic metasurfaces. Nonlinear metalenses with the focal lengths of 30 μm, 50 μm and 100 μm are demonstrated experimentally, providing a direct evidence for both generating and manipulating SH signals based on the nonlinear hybrid metasurfaces. It shows great potential applications in designing of integrated, ultra-thin, compacted and efficient nonlinear optical devices, such as frequency converters, nonlinear holography and generation of nonlinear optical vortex beam.

Optical metasurfaces are a class of artificial nanostructured surfaces with the engineered linear optical properties, providing an ideal platform for designing integrated, ultra-thin and compacted planer linear optical devices.[1-9] Recently, this concept has been extended to the nonlinear regime where nonlinear optical metasurfaces are able to fully control the wavefront of generated nonlinear signals.[10-16] However, this new paradigm faces important challenges since it is required to act as an efficient generator and a sub-wavelength phase-modulator of nonlinear signals simultaneously.

Second-harmonic generation (SHG) is an important second-order nonlinear process that generates photon with twice the fundamental frequency of the pumping laser.[17] Generally, SHG in metal-based (plasmonic) metasurfaces have attracted extensive attentions for applications in the nonlinear regime since plasmonic metasurfaces have an impressive performance in the linear regime with a reliable fabrication technology.[18,19] However, it suffers from a low SHG conversion efficiency due to the inherently weak nonlinear response ($\chi^{(2)}\sim$ pm/V).[18,20] Although it can be boosted by orders of magnitude at resonance condition, the strong absorption losses lead to damage by accumulated heat effect, especially in the visible region. Meanwhile, it might be interfered by the unwanted signals from higher-order nonlinearity, such as white-light supercontinuum (WLC).[21,22] Therefore only a very low pumping intensity is allowed, showing limitations for practical applications. So far, compared with a large number of works on SHG in metal nanostructures, only a few works have demonstrated the engineered SH signals by plasmonic metasurfaces[11-13,23], as well as third-harmonic generation (THG)[12,15] and four-wave mixing (FWM)[14].

Alternatively, a scheme of nonlinear hybrid metasurfaces has been proposed by coupling the plasmonic metasurfaces to a highly efficient nonlinear semiconductor. M. A. Belkin and his co-workers have demonstrated a remarkably large SHG susceptibility ($\chi^{(2)}\sim$54 nm/V) in the mid-infrared region (at ~8 μm) by a plasmonic metasurface coupled to multi-quantum-well (MQW) semiconductor structures.[24,25] Although this hybrid metasurface is suitable for the mid-infrared region, it is unlikely to work in the visible region. The available MQW structures are obviously too thick (~400 nm) for a femtosecond laser excitation at 810 nm (~$\lambda$/2), leading to the problem that the emitted SH signal may be out of phase during its propagation in the MQW structures. More importantly, the intrinsically large nonlinear susceptibility of MQW is $\chi_{zzz}^{(2)}$ that is polarized in the direction normal to the surface. It implies that it is not likely to achieve such a large nonlinear susceptibility with an ultrathin MQW (<40 nm, ~$\lambda$/20 for ~810 nm). Therefore it is still quite necessary to demonstrate a new type of nonlinear hybrid metasurface in the visible region. Recently, new types of atomically thin crystals based on layered materials, such as transition metal dichalcogenides (TMD), exhibits excellent physical and chemical properties.[26-28] Specifically, the TMD family of monolayer materials is ultrathin with only one-atom thickness and can be easily transferred to the metal nanostructures in a simple approach.[29,30] In addition, characterization of $WS_2$ and $MoS_2$ monolayers has also revealed a large intrinsic nonlinear susceptibility, as high as ~nm/V in the visible region.[31,32] Thus, it provides an elegant and promising way to overcome these constraints.

Here we propose and experimentally demonstrate a new type of nonlinear hybrid metasurfaces in the visible region. It consists of a $WS_2$ monolayer and a phased gold nanohole array, which is easily prepared by a material transferring method. The results of SHG measurement indicate that a large SHG susceptibility of ~$10^{-1}$ nm/V at 810 nm is achieved, which is 2~3 orders of magnitude larger than that of typical plasmonic metasurfaces. The nonlinear metalenses with the focal lengths of 30 μm, 50 μm and 100 μm are demonstrated experimentally, and the small focal spots of the metalenses are achieved (1~2 μm in diameter). It provides a direct evidence both for efficiently generating and manipulating SH signals based on this nonlinear hybrid metasurfaces, which shows the great potentials to integrated, ultra-thin, compacted and efficient nonlinear optical devices.

Figure 1(a) shows a brief illustration of the fabrication process of $WS_2$-gold nanohole hybrid metasurfaces. Firstly, polymethyl methacrylate (PMMA) in the anisole solution was drop-coated onto WS$_2$ flakes supported by a sapphire substrate and baked to remove the solvents. The sample was then immersed in NaOH solution to etch the sapphire substrate and the PMMA-WS$_2$ film lifted off from the growth substrate. The PMMA-WS$_2$ film was fished out of the NaOH solution by a glass slide and immersed again in the deionized water, offering a floating PMMA-WS$_2$ film on the surface of the deionized water. Meanwhile, a gold film was deposited on a quartz substrate and milled with nano-patterns by focused ion beams (FIB) method. The important step is that the PMMA-WS$_2$ films were fished out by the fabricated metasurfaces and aligned quickly with the gold nanohole array under an optical microscope. Finally, it was washed in acetone to remove the PMMA coating.

The WS$_2$ flake grown on the sapphire substrate and the transferred WS$_2$ flake on gold film are shown in Figure 1(b) and (c) respectively. It can be seen clearly that the triangular WS$_2$ flakes remain intact after transferring process. The WS$_2$ flakes used in the experiment is confirmed to be a monolayer with a thickness of 0.7 nm by atomic force microscope (AFM) measurement shown in the inset of Figure 1(b). Figure 1(d) exhibits an optical microscope image of the hybrid metasurface, indicating that the gold nanohole array is fully covered by the WS$_2$ monolayer. The details of morphology are further characterized by scanning electron microscopy (SEM), shown in Figure 1(e). Although the WS$_2$ monolayer on the gold nanoholes array is in good condition, some tiny cracks are observed in the WS$_2$ monolayer. It might be induced by the stress relief when the PMMA film was removed. It should be noted that the tiny cracks have a negligible influence on the performance of the hybrid metasurfaces according to the experimental results described below.

A nonlinear metalens was designed for a demonstration of this nonlinear hybrid metasurfaces. Similar to the planar lenses by linear metasurfaces[33,34], it is crucial that a well-designed phase profile must be imparted to the emitters of nonlinear signals for focusing. In our experiment, it can easily be realized by taking the gold nanoholes array as a phase plate. As shown in Figure 2(a), an orientation angle, $\theta$, is defined as the angle between the long-axis of the rectangular gold nanohole and *x*-axis in the labor frame. When a left (right) circular polarized fundamental beam passes through the rectangular gold nanohole, the transmitted fundamental beam with a right (left) circular polarization is delivered by a geometric phase delay of $2\theta$ while keeping the beam amplitude as the same.[34,35] When it further excites the WS$_2$ monolayer, the emitted SH signal will be carried a phase delay of $4\theta$ since SHG is a second-order nonlinear process. Therefore, the phase of the emitted SH signal can be modulated by designing the rectangular gold nanohole with different orientations angle, $\theta$. To act like a spherical lens, the phase profile $\varphi(x, y)$ of the nonlinear metalens needs to follow by,

$$\varphi(x,y) = 2\pi(f - \sqrt{x^2 + y^2 + f^2})/\lambda + 2n\pi , \quad (1)$$

where $\lambda$ is the design wavelength, $n$ is an arbitrary integer, $x$ and $y$ are the coordinates of each nanohole, and $f$ is the focal length. Here we chose $\lambda$ to be the SH wavelength at 405 nm and $f$ is equal to 50 μm. And then, the phase profile $2\theta(x, y)$ $(=\varphi(x, y)/2)$ that acts on the WS$_2$ monolayer is obtained as shown in Figure 2(b). Figure 2(c) exhibits the fabricated rectangular nanoholes array with varied orientations according to the phase profile, and the details of the metasurface is shown in Figure 2(d). The total diameter of the fabricated metasurface was 20 μm. Subsequently, a WS$_2$ monolayer was transferred on the metasurface by using the method described in Figure 1(a).

Figure 3(a) illustrates a scheme of a home-built microscope system at room temperature for the SHG measurements. A circular polarized laser was focused by an 8 cm lens, and then it excited the WS$_2$ monolayer after it passed through the gold nanoholes array. The generated signal was either collected by a spectrometer through a fiber or directly entered into a CCD camera. In Figure 3(b), the red curve shows the spectrum of the emitted signal of the hybrid metasurface. It is observed that a sharp peak is located at 405 nm, which is exactly the twice frequency of the pumping laser at 810 nm. Moreover, Figure 3(d) presents the measured signal intensity as a function of the square of the pumping power $P^2$. One can clearly see that the intensity increases linearly with $P^2$. Thus, it is concluded that the measured signal is ascribed to the SHG response in our experiment.

For comparison, the SH signal of the same bare metasurfaces without a $WS_2$ monolayer was also measured under the same experimental condition, which is indicated by the black curve in Figure 3(b). Obviously, the SHG intensity of the hybrid metasurface is much larger than that of the same bare metasurface, indicating a significant SHG enhancement. For a quantitative estimation of the enhancement, we enlarged the spectrum of the bare metasurfaces shown in Figure 3(c). Since it was submerged in background noises, the intensity of SH signal of the bare metasurfaces was estimated to be less than 6 arbitrary units (a.u.) reasonably. A SHG enhancement factor is confirmed to be at least 2 orders of magnitude. Furthermore, we also compared the SHG intensity of the hybrid metasurfaces with that of bare $WS_2$ monolayer quantitatively, and it indicates a 10 times decrease of SHG conversion efficiency. It has been reported that the $WS_2$ monolayer possesses a large second-order susceptibility in the level of ~nm/V,[32] and then the second-order susceptibility of our hybrid metasurfaces is at ~$10^{-1}$ nm/V. Therefore, the second-order susceptibility of our hybrid metasurfaces is 2~3 orders of magnitude larger than that of typical plasmonic metasurfaces.

The focusing of the nonlinear metalens based on the hybrid metasurface was determined subsequently. Figure 4(a) presents the spatial intensity profile of emitted SH signals from the hybrid metasurface located in the center of plane at $z=0$ μm. An evident converging of SHG along $z$-axis can be observed, leading to a bright focal spot of SHG. The focal length is measured to be 49 μm, which is consistent with the designing value of 50 μm. Figure 4(b) shows the intensity distribution of SHG along $x$-axis at the focal plane ($y=0$ μm, $z=49$ μm). The diameter of the SHG focal spot was measured to be ~1.6 μm, while the inset of Figure 4(b) presents an optical image captured at the focal plane ($z=49$ μm). Furthermore, the intensity distribution of SHG along $z$-axis ($x=0$ μm, $y=0$ μm) across the focal spot is shown in Figure 4(c). It is worth noting that the phase profiles of the hybrid metasurface for the fundamental beam and SH emission are $\varphi(x, y)/2$ and $\varphi(x, y)$, respectively. Thus, the transmitted fundamental beam and SH emission can be separated spatially.

It is convenient to control the focusing behavior by changing the phase profile of nonlinear metalens. We also fabricated nonlinear metalenses with the focal lengths of 30 μm and 100 μm. According to equation (1), the phase profiles $2\theta(x, y)$ $(=\varphi(x, y)/2)$ that acts on the $WS_2$ monolayer are acquired for focal lengths of 30 μm and 100 μm, which is shown in Figure 5(a) and 5(c) respectively. Figure 5(d) and 5(f) present the relevant results of the measured nonlinear focusing at $f$=30 μm and 100 μm, respectively. Meanwhile, the result for the nonlinear metalens at $f$=50 μm is also listed in Figure 5(b) and 5(e) for comparison. The diameters of the three focal spots are measured to be in scale of ~1 μm, which is consistent with the result in previous works. Moreover, we calculated the 2D linear focusing of the three metalenses with the same phase profile. It agrees with the experimental results well, which further supports the nonlinear metalens based on our hybrid metasurface. Therefore, it can be concluded that the hybrid-metasurface based nonlinear metalenses provide an efficient generation of SH signals and effective focusing simultaneously. The arbitrary focal length of the nonlinear metalens can be obtained by a proper design of the metallic nanoholes distributions.

It is worth noting that plasmonics nanohole is suitable for the nonlinear hybrid metasurfaces in our experiment rather than plasmonic nanoatennas. Specifically, it enables "tailoring" of the $WS_2$ monolayer for controlling SHG because only the fundamental beam that passes through nanoholes can excite the $WS_2$ monolayer, while the rest is blocked by the unperforated films. This significantly reduces the background noise of SHG. The main drawback of the nanohole is a relatively low transmittance of the fundamental beam, leading to a decrease of the effective SHG conversion efficiency in our hybrid metasurface (~10%). However, it can be significantly improved by changing the phase control method. It has been reported that the cross-polarization coupling efficiency is fundamentally limited to stay below 25% for geometric phase[36], and it can be increased to ~100% by adopting alternative phase control method of tailoring the nanostructures size and shape[37-39]. Furthermore, a higher SHG conversion efficiency can be achieved by using low-loss dielectric materials for reducing of the absorption to fundamental beam and SH signals.[40,41] Although only nonlinear focusing is presented in this work, this new type of nonlinear hybrid metasurface is able to realize versatile nonlinear signals manipulation by properly designing the phase profiles of the plasmonic metasurface, such as nonlinear beam direction manipulation, nonlinear vortex beam generation and nonlinear holography. In addition, a polarization control of the SH emission has also been proved by choosing the hybrid metasurface with constituent gold nanoholes of threefold rotational symmetry.

In conclusion, we propose and experimentally demonstrate a new type of nonlinear hybrid metasurfaces for nonlinear transformation optics in the visible region. It consists of a gold-nanohole based plasmonic metasurfaces coupled to a monolayer of $WS_2$, which can be easily prepared by a material transferring method. A large SHG susceptibility of ~$10^{-1}$ nm/V at 810 nm is achieved, which is 2~3 orders of magnitude larger than that of typical plasmonic metasurfaces. Furthermore, the hybrid metasurfaces effectively suppress any unwanted signals from higher-order nonlinearity (WLC) even under a high pumping intensity, which is proved to be more stable and reliable under an intensive pumping. For the capability of phase modulation, the nonlinear metalenses with the focal lengths of 30 μm, 50 μm and 100 μm are demonstrated experimentally. It provides a direct evidence for both generating and manipulating SHG signals based on this nonlinear hybrid metasurfaces. From the point of application view, it enables a miniaturized nonlinear optical device with the good performance in a small scale of tens of nanometers. Therefore, it provides a wide range of opportunities for the creation of integrated, ultrathin, ultra-compact nonlinear optical devices, such as frequency converters, all-optical switches and the generation of optical vortex beam, as well as nonlinear holography.

## Methods

Sample fabrication. (1) The lift-off of WS$_2$ monolayer. First, 1 g of polymethyl methacrylate (PMMA, Aldrich, average MW 996 000) was dissolved in anisole to prepare 50 g of solution (2 wt%). A few drops of this solution were drop-coated on the sapphire substrate with WS$_2$ monolayer flakes until it covered all the substrate. The substrate was then placed at room temperature for 1.5 h for the evaporation of the anisole, followed by curing step of baking at 120 °C for 0.5 h. About 1 mm wide polymer strips were scratched by a blade at the edges of the substrate to reduce the lift-off time. The substrate was immersed in 3 mol/L NaOH aqueous solution at 50 °C for more than 4h to etch the sapphire surface, and the PMMA-WS$_2$ films would lift off from the growth substrate naturally. The films were fished out with a glass slide and immersed in deionized water for 3 times in order to remove the residual NaOH solution. (2) The fabrication of gold metasurfaces. Using an e-beam evaporator, we deposited a 5-nm thick Cr adhesion layer and a 60-nm thick gold layer on a quartz substrate. The different arrangement of nanoholes were milled on the film by means of focused ion beam milling method (FIB, FEI Versa 3D) to generate the metallic metasurfaces. (3) The combination of the WS$_2$ monolayers with the gold metasurfaces. With the help of tweezers, the PMMA-WS$_2$ films were fished out with the fabricated metasurfaces and aligned quickly with the metasurface arrays under the microscope before the water dried. The substrate was then baked at 120 °C for 1 h to improve the combination of WS$_2$ monolayers with the gold metasurfaces. Finally, the substate was immersed in acetone for 3 times (each time 15 mins) to remove the PMMA film.

Experimental set-up and measurement of SH signals. A mode-locked Ti-sapphire femtosecond laser centered at 810 nm (Vitara Coherent, 8 fs and 80 MHz) was used for the fundamental beam source. It was filtered spectrally to avoid residual SH from the laser, and its polarization was controlled by a quarter-wave plate. The emission from the sample was collected with an objective lens (Olympus, 40 × and 0.65NA), filtered spectrally, and directly imported to a CCD or through a fiber to a spectrometer (Princeton Instruments Acton 2500i with Pixis CCD camera). The emitted SH signal was extracted by a Glan-laser polarizer with a quarter-wave plate at the SH wavelength.

<u>Measurement of focusing effect.</u> A home-built 3D imaging system was used to measure the emitted SH signal profiles from the hybrid metasurfaces under an excitation of a plane-wave laser. The laser was focused by an 8 cm lens to generate the collimated incident fundamental beam. Quarter-wave plates and polarizers were used to generate the circular polarized incident fundamental beam and select SH signals with an opposite circular polarization. To measure the focusing of SHG, we captured the SHG images at the different planes from 0 μm to 150 μm along $z$-axis with a step-size of 0.5 μm. The transmitted SH signal was collected by an objective (40×/0.65) and imaged on a CCD camera.

## Acknowledgment

This work was supported by the 973 Programs under grants 2014CB921301 and National Natural Science Foundation of China (Nos. 11204097 and 11674117), the Doctoral fund of Ministry of Education of China under Grant No. 20130142110078. Special thanks to the Center of Nano-Science and Technology of Wuhan University for their support of sample fabrication.

## Author contributions

K. W. and P. X. L. conceived the experiments; J. W. C and W. W. L. performed the sample fabrication; J. W. C, H. B. H. and X. B. H. performed optical experiments and data analysis; B. W. and H. L. performed theoretical calculations; K. W. and J. W. C. wrote the manuscript.

## Competing financial interests

The authors declare no competing financial interests.

# Figure Captions

**Figure 1** The fabrication of $WS_2$-gold hybrid metasurfaces. (a) Schematic for transferring $WS_2$ monolayers onto gold nanohole metasurfaces by the polymethyl methacrylate (PMMA)-assisted transfer method. (b) and (c) The optical images of a $WS_2$ monolayer before (on sapphire substrate) and after (on gold film) transferring, respectively. The height profile of the $WS_2$ monolayer was measured by AFM, from which a height of 0.7 nm can be extracted. Scale bars: 20 μm. (d) and (e) The optical and SEM images of $WS_2$ on gold nanohole arrays. Scale bars: (d) 10 μm, (e) 2 μm.

**Figure 2** The design and fabrication of the $WS_2$-gold nanohole hybrid metasurfaces for nonlinear focusing. (a) The geometry of the designed rectangular nanohole unit cell structure in the metasurface, with periodicity $P_x = P_y = 400$ nm. The thickness of gold film is $H = 60$ nm. The nanoslit has length $L = 185$ nm and width $W = 80$ nm, and it can rotate in the *x-y* plane with an orientation angle, $\theta$. All sharp corners of the nanoholes are smoothened out by the cylindrical surfaces with a radius of 15 nm. (b) The phase distribution that acts on the $WS_2$ monolayer for nonlinear focusing with a focal length of 50 μm. (c) The SEM image of the fabricated gold metasurface with rectangular nanoholes of different orientations, $\theta$. The details in the rectangular area are shown in (d). Scale bars: (c) 5 μm, (d) 1 μm.

**Figure 3** The nonlinear characteristics of hybrid metasurfaces with/without the $WS_2$ monolayer. (a) The schematic sketch of the experiment setup for SHG and focusing measurements. H1: half-wave plates; H2: Glan-laser polarizer; Q: quarter-wave plate; M1, M2: mirrors; LP: 600-nm long pass filter; SP: 650-nm short pass filter; BP: 400-nm band pass filter. (b) The nonlinear emission spectra of rectangular nanohole-$WS_2$ hybrid metasurfaces (w/ $WS_2$) and the bare rectangular nanoholes (w/o $WS_2$) under the excitation of the circularly polarized fundamental beam. (c) The enlarged emission spectrum of the bare rectangular nanoholes. (d) The measured SHG intensity at 405 nm as a function of the square of incident laser power $P^2$.

**Figure 4** The nonlinear focusing characteristics of the WS$_2$-gold nanohole hybrid metasurfaces. (a) The experimental results of SHG focusing by using the hybrid metasurfaces. (b) and (c) The SHG intensity distribution along *x*-axis and *z*-axis across the center of the focal spot, respectively. The inset of (b) shows the SH beam *x-y* cross sections at the focus of 49 μm.

**Figure 5** Nonlinear metalenses of focal lengths of 30, 50 and 100 μm based on the WS$_2$-gold nanohole hybrid metasurfaces. (a–c) The phase distribution that acts on the WS$_2$ monolayer for nonlinear focusing. (d–f) The metalens experimental measurement of the focal region, respectively.

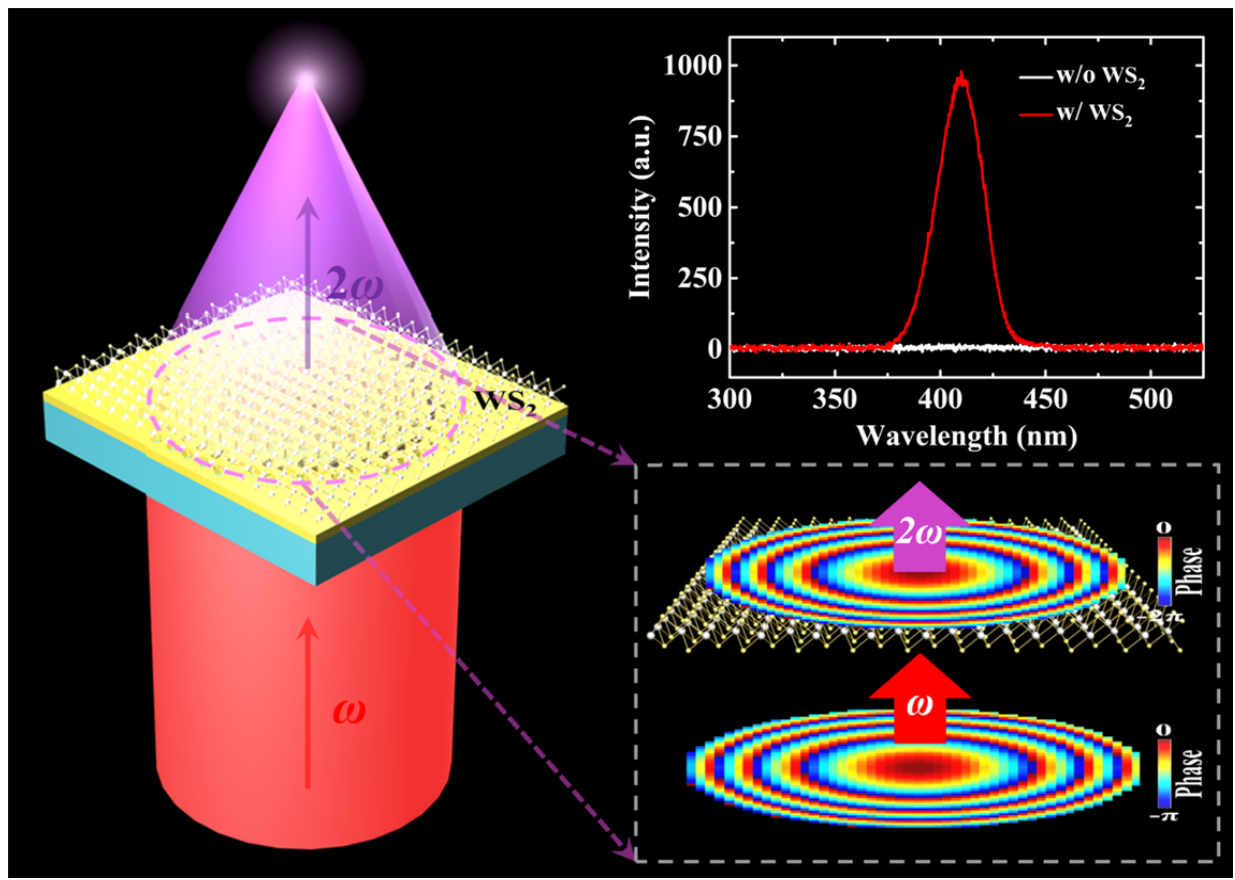

**TOC**

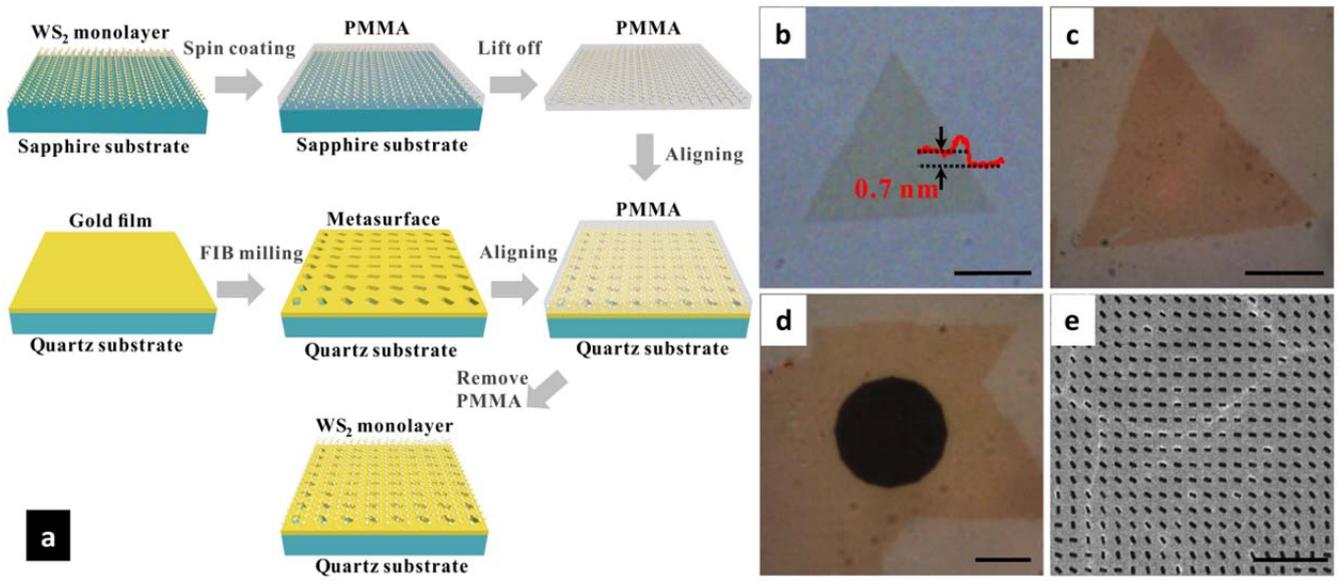

**Figure 1**

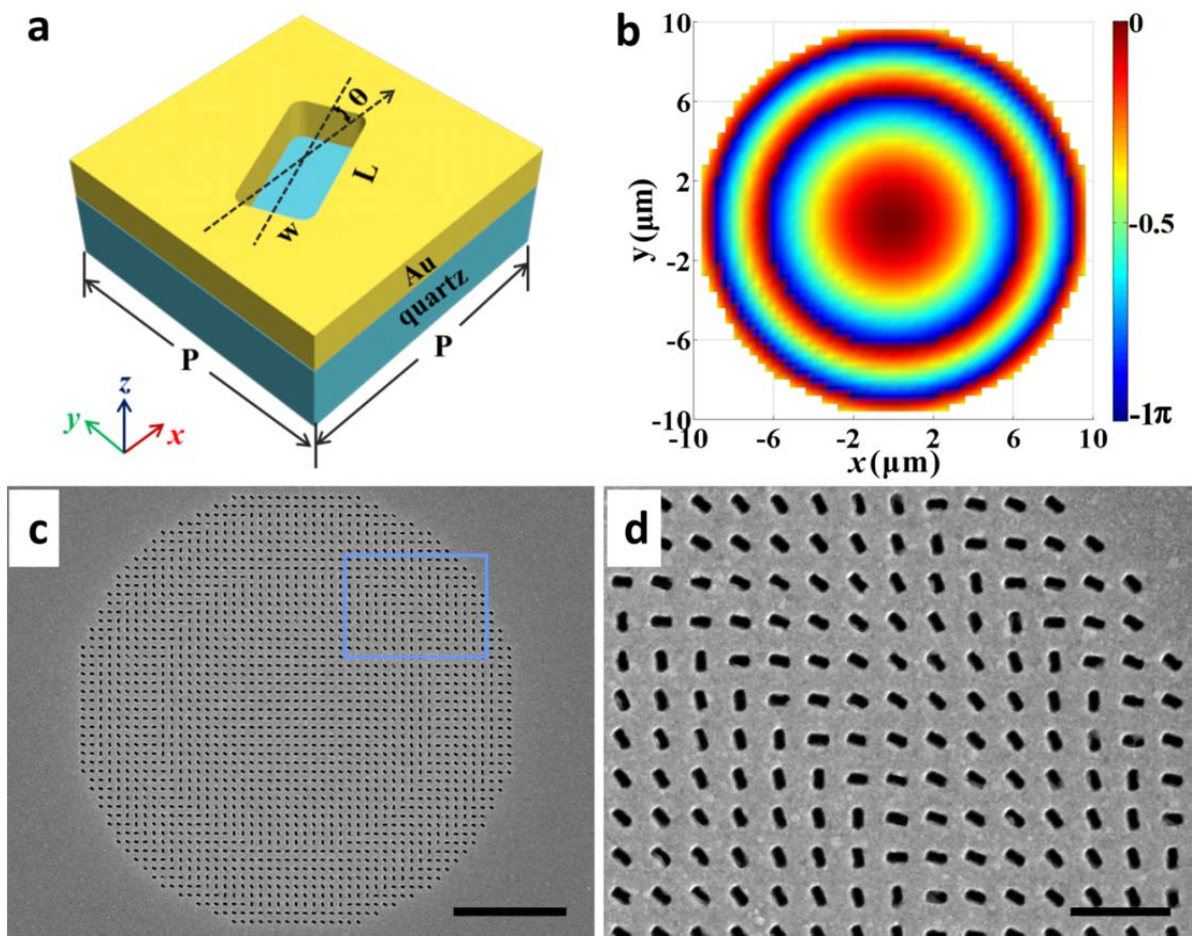

**Figure 2**

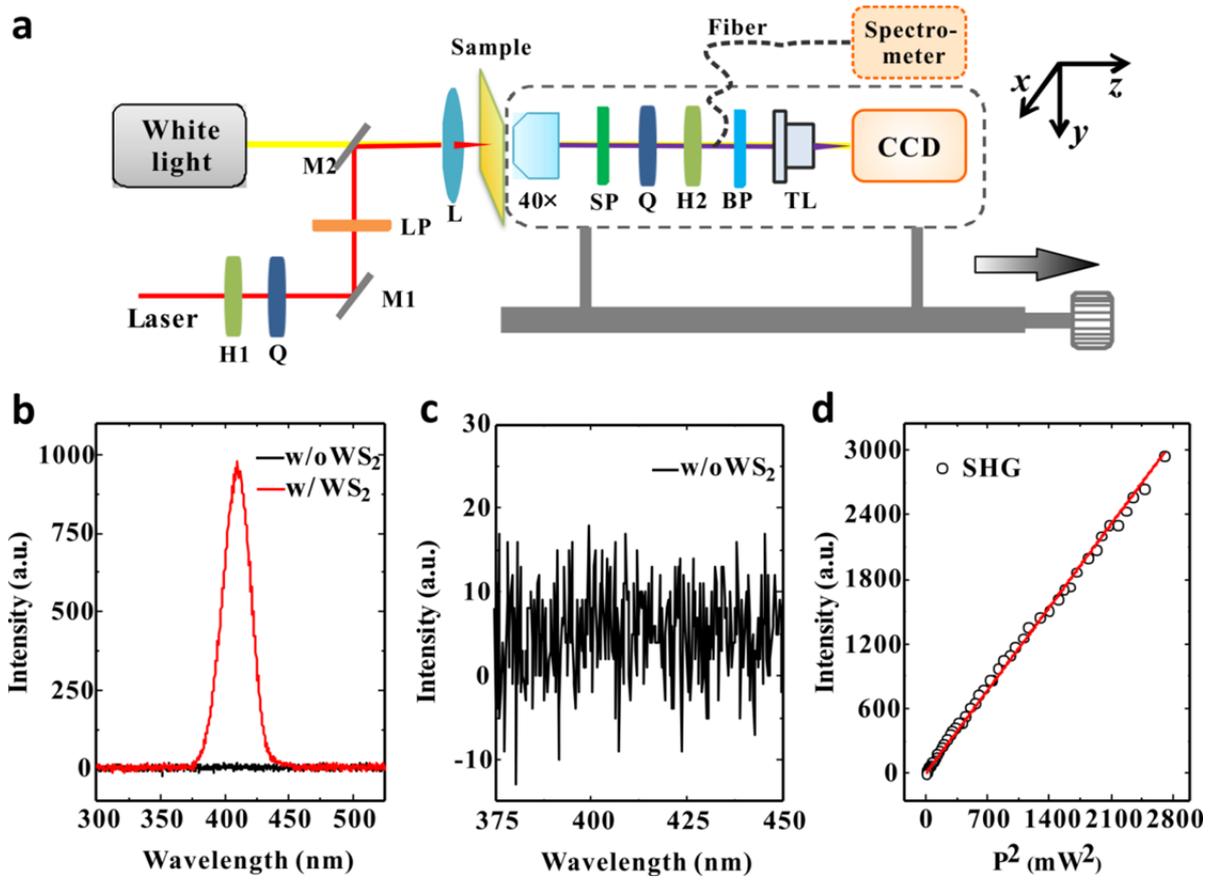

**Figure 3**

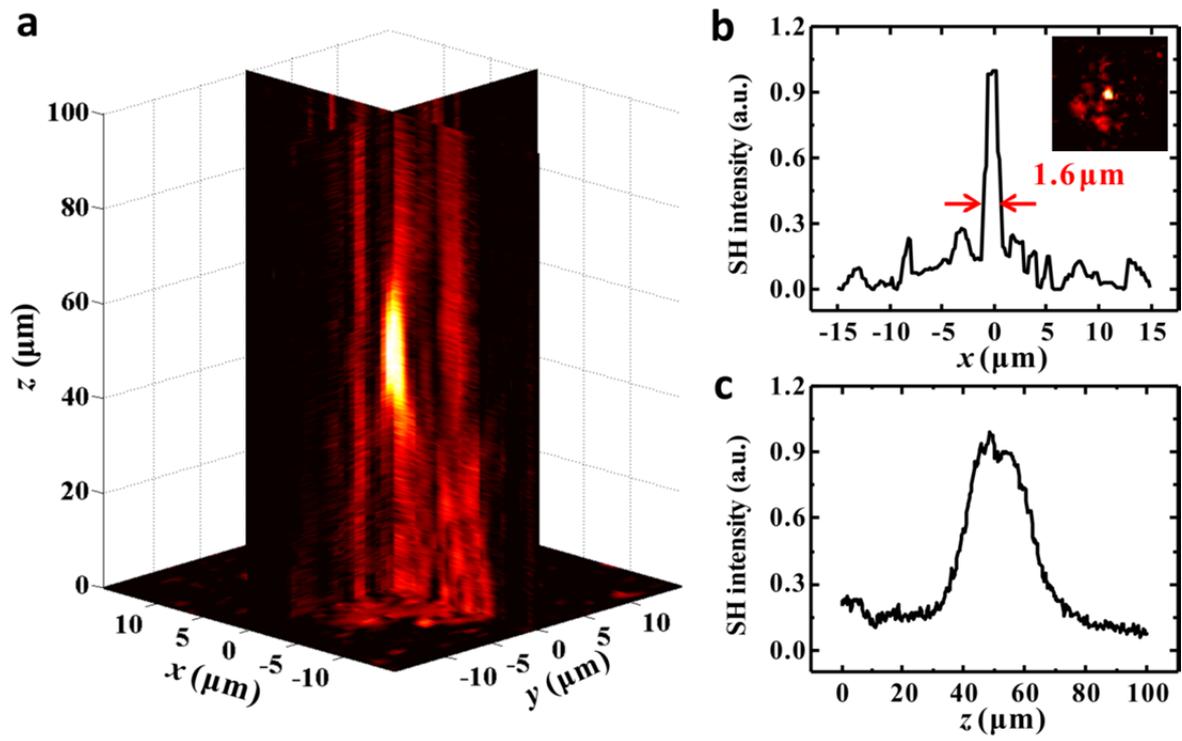

**Figure 4**

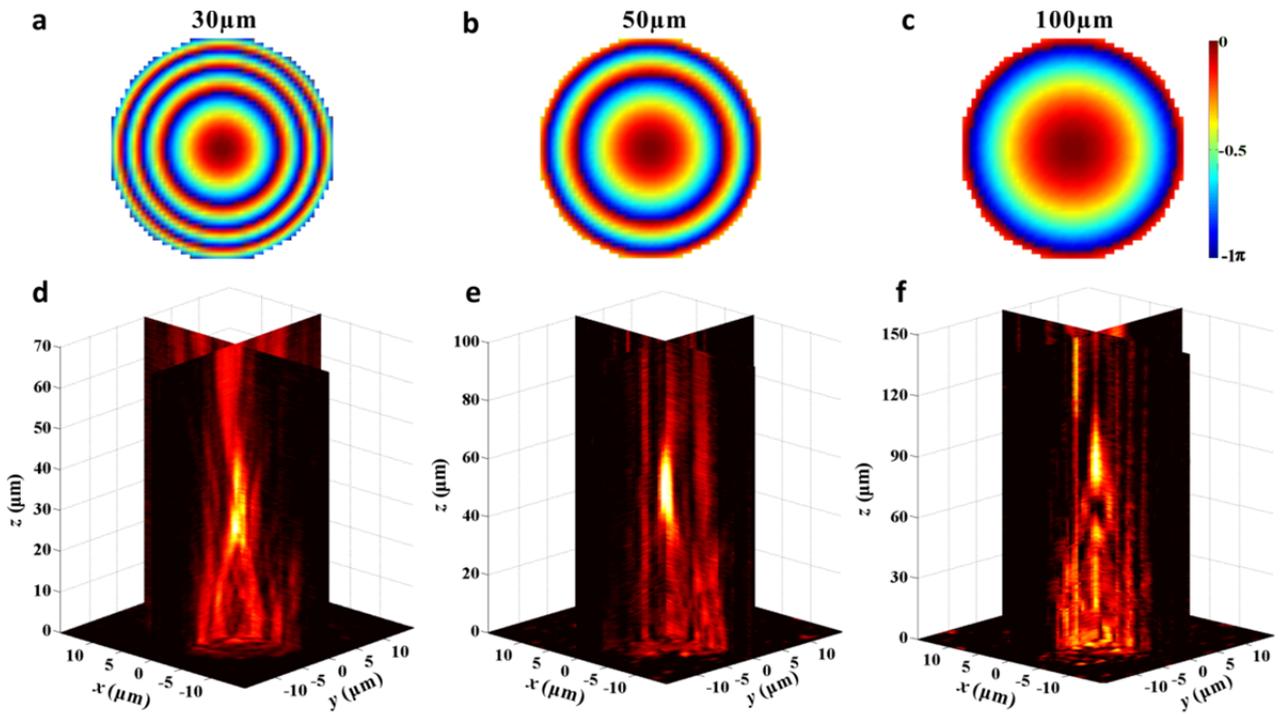

**Figure 5**